\documentclass[11pt,a4paper]{article}
\usepackage{amsmath}
\usepackage[mathcal]{eucal}
\usepackage{latexsym}
\usepackage{amssymb}
\begin{titlepage} 
\title{Instanton representation of Plebanski gravity: a conceptual introduction}
\author{Eyo Eyo Ita III}
\medskip  
\input amssym.def
\input amssym.tex

\def \in{\indent}

\begin{document}  
\maketitle
\bigskip
\centerline{Physics Department, US Naval Academy} 
\smallskip
\centerline{Annapolis, Maryland}
\smallskip
\centerline{ita@usna.edu}
           
\bigskip    
                  
\begin{abstract}
This paper presents a self-contained introduction into the instanton representation of Plebanski gravity (IRPG), a GR formulation which uses a gauge connection and a three by three matrix as basic variables.  In this paper we explore some physical consequences of the IRPG both from the 3+1 and from the covariant perspectives, including its regime of equivalence with Plebanski's theory, and gravitational instantons.  We conclude with the IRPG evolution equations, which are shown to preserve the initial value constraints of the theory. 
\end{abstract}
\end{titlepage}

\section{Introduction}

The idea of gravity as a theory where the spacetime metric loses its privileged status as a fundamental variable has had a long history.  In the 1980's the Ashtekar formulation was introduced (see e.g. \cite{ASH1}, \cite{ASH2}, \cite{ASH3}), which rewrote General Relativity in terms of a set of Yang--Mills-like variables known as the Ashtekar variables.  The Ashtekar formulation, as initially presented in 3+1 form, proved to be most conducive for nonperturbative quantization techniques via the Hamiltonian formalism.  From this formalism arose the loop quantization programme \cite{REPORT}.  The Ashtekar formulation was first re-written in a manifestly covariant form by Jacobson and Smolin \cite{SAMUEL1}, \cite{SMOLJAC}.  This spurned further developments for example as in \cite{SPINCON} and \cite{SELFDUAL}, where Capovilla, Dell and Jacobson (CDJ) found a covariant version of Ashtekar's gravity almost entirely in terms of a spin connection.  These authors (CDJ) independently discovered a covariant gravity formulation where self-dual two forms replace the spacetime metric \cite{SELFDUAL}.  Incidentally, the equations of motion arising from this latter formulation were written down and analyzed in the 1970's by Jerzy Plebanski \cite{PLEBANSKI}.  It was realized by CDJ that the Ashtekar formulation is none other than a 3+1 form of Plebanski's action.  The aforementioned developments signify a shift from the Hamiltonian-type 3+1 formalism to more covariant approaches for gravity.  Indeed, the Plebanski formalism can be seen as a kind of mother theory for these formulations.\par 
\indent 
The covariant form of a gravity theory is asthetically pleasing from various perspectives, including perturbation theory.  The 3+1 Hamiltonian form, while not manifestly covariant, plays an important role in addressing the evolutionary aspects of the classical equations, and for the quantization of totally constrained systems such as gravity.  So both forms are important, and provide complementary strengths and insights.  In this paper we will present a formulation of gravity which we have named the instanton representation of Plebanski gravity (IRPG), for reasons which will become clear.  We will present the theory initially in a 3+1 form, using a different combination of basic variables than the aforementioned formulations, where again the spacetime metric is not fundamental but becomes a derived quantity.  We will examine the IRPG equations from the 3+1 as well as from the covariant perspectives, highlighting the importance of these two perspectives in complementing each other.  We will ultimately show for certain regimes that the IRPG is really Plebanski's theory in disguise.  A ramification of this is that the IRPG, in addition to the Ashtekar formulation, can be included in the list of 3+1 formulations of Plebanski's gravity.\par 
\indent 
This paper is organized as follows.  In Section 2 we present the IRPG in 3+1 form, and write down the constraint equations.  Sections 3 and 4 cover the equations for the dynamical variables $A$ and $\Psi$, where we provide a physical interpretation for the covariant meaning of these equations.  In section 5 we recapitulate the previous results into overall context, demonstrating the equivalence between the IRPG and the Plebanski actions.  In this section we also obtain the evolution equations for $A$ and $\Psi$ from a 3+1 perspective independently of any symplectic structure.  Additionally, we prove that the constraint equations for the IRPG are preserved under time evolution by the evolution equations.  Section 7 is a summary of the paper, including directions for future research.  Usually, in the introduction of some new formalism for gravity, there should be some sort of rationale for why one would want to use it in lieu of existing approaches: in short, what aspects of the IRPG could make it useful.  While we will expose some interesting relationships throughout the body of this paper, the ultimate motivations will become more apparent once the final form of the equations of motion have been displayed all in one place.\footnote{It amounts to the observation that there appear to be fewer spatial derivatives acting on certain auxiliary fields which one needs to worry about in the IRPG in relation to other approaches.  This feature could facilitate gauge-fixing procedures as well as the construction of General Relativity solutions in practice.}

\section{The starting action}

The action for the instanton representation of Plebanski gravity is given in 3+1 form by
\begin{eqnarray} 
\label{INSTANTONREP}
I_{Inst}=\int{dt}\int_{\Sigma}d^3x\biggl[\Psi_{ae}B^i_e\dot{A}^a_i+A^a_0B^i_eD_i\Psi_{ae}\nonumber\\
-\epsilon_{ijk}N^iB^j_eB^k_a\Psi_{ae}+\beta{N}\sqrt{\hbox{det}B}\sqrt{\hbox{det}\Psi}\bigl(\Lambda+\hbox{tr}\Psi^{-1}\bigr)\biggr],
\end{eqnarray}
\noindent 
where $\Sigma$ is a 3-dimensional spatial manifold of a given topology embedded in a spacetime of topology $M=\Sigma\times{R}$.  The basic dynamical fields for $I_{Inst}$ are a $SO(3)$ gauge connection $A^a_{\mu}$ and matrix $\Psi_{ae}\in{SO}(3)\otimes{SO}(3)$,\footnote{For index conventions, internal $SO(3)$ indices are labelled by symbols from the beginning of the Latin alphabet $a,b,c,\dots$, and spatial indices $i,j,k$ from the middle.  Greek symbols $\mu,\nu,\dots$ will denote 4-dimensional spacetime indices.} where
\begin{eqnarray} 
\label{NONDEGENERATE}
(\hbox{det}\Psi)\neq{0};~~(\hbox{det}B)\neq{0}.
\end{eqnarray} 
\noindent 
The connection $A^a_{\mu}$ splits into temporal and spatial parts $A^a_0,A^a_i$, and its field strength 
$F^a_{\mu\nu}=\partial_{\mu}A^a_{\nu}-\partial_{\nu}A^a_{\mu}+f^{abc}A^b_{\mu}A^c_{\nu}$ splits into temporal part containing velocities $\dot{A}^a_i$, and a spatial part $B^i_a$ known as the magnetic field 
\begin{eqnarray} 
\label{CURVE}
F^a_{0i}=\dot{A}^a_i-D_iA^a_0;~~B^i_a={1 \over 2}\epsilon^{ijk}F^a_{jk}.
\end{eqnarray}
\noindent 
The fields $N$ and $N^i$ are auxiliary fields and $\beta$ is a numerical constant which depends on the signature of spacetime.\footnote{For Lorentzian signature we have $\beta=\pm{i}$, and for Euclidean signature $\beta=\pm{1}$.}  Additionally, the operator $D_{\mu}$ is the $SO(3)$ covariant derivative, which acts as
\begin{eqnarray} 
\label{COVARIANTDERIVATIVE}
D_{\mu}V_a=\partial_{\mu}V_a+f_{abc}A^b_{\mu}V_c;\nonumber\\
D_{\mu}m_{ae}=\partial_{\mu}m_{ae}+A^b_{\mu}\bigl(f_{abc}m_{ce}+f_{ebc}m_{ae}\bigr)
\end{eqnarray} 
\noindent 
on $SO(3)$-valued 3-vectors $V_c$ and second-rank tensors $m_{ae}$.\par 
\indent 
The quantities $N^i$, $N$ and $A^a_0$ are auxiliary fields since their velocities do not appear in the action (\ref{INSTANTONREP}), but $\Psi_{ae}$ is not an auxiliary field.  While the velocty of $\Psi_{ae}$ does not explicitly appear, it is clear from (\ref{INSTANTONREP}) that $\Psi_{ae}$ is a dynamical variable on the same footing as $A^a_i$, since a $\dot{\Psi}_{ae}$ term can be induced through an integration by parts, which is not the case for $(N,N^i,A^a_0)$.  The equations of motion for the auxiliary fields $N^i$, $N$ and $A^a_0$ are given by
\begin{eqnarray} 
\label{CURVEFIT}
{{\delta{I}_{Inst}} \over {\delta{N}^i}}=\epsilon_{ijk}B^j_eB^k_a\Psi_{ae}=(\hbox{det}B)(B^{-1})^d_i\psi_d=0;\nonumber\\
{{\delta{I}_{Inst}} \over {\delta{N}}}=\beta\sqrt{\hbox{det}B}\sqrt{\hbox{det}\Psi}\bigl(\Lambda+\hbox{tr}\Psi^{-1}\bigr)=0;\nonumber\\
{{\delta{I}_{Inst}} \over {\delta{A}^a_0}}\equiv{G}_a=B^i_eD_i\Psi_{ae}=0,
\end{eqnarray}
\noindent 
where we have defined 
\begin{eqnarray} 
\label{CURVE2}
\psi_d=\epsilon_{dae}\Psi_{ae}
\end{eqnarray} 
\noindent 
as the antisymmetric part of $\Psi_{ae}$.  Equations (\ref{CURVEFIT}) are constraint, not evolution, equations since they do not have any time derivatives.  Due to the nondegeneracy condition (\ref{NONDEGENERATE}), equations (\ref{CURVEFIT}) are equivalent to
\begin{eqnarray} 
\label{CURVE1}
\psi_d=0;~~\Lambda+\hbox{tr}\Psi^{-1}=0;~~B^i_eD_i\Psi_{ae}=0.
\end{eqnarray} 
In the next two sections we will find the Lagrange's equations for the dynamical fields $\Psi_{ae}$ and $A^a_{\mu}$, and examine their physical ramifications.\par 
\indent 

\section{Lagrange's Equation for $\Psi_{ae}$}

To find the Lagrange's equation for $\Psi_{ae}$ it will be convenient to integrate (\ref{INSTANTONREP}) by parts, discarding boundary terms.\footnote{For compact manifolds this procedure is inherently justified.  For the noncompact case we must assume that the fields fall off sufficiently rapidly at infinity.  The latter violates the nondegeneracy conditions (\ref{NONDEGENERATE}) at infinity, though not within the bulk of spacetime.}  This yields the following action
\begin{eqnarray}
\label{CANON9}
I_{Inst}=\int{dt}\int_{\Sigma}d^3x\biggl[\Psi_{ae}B^i_e\bigl(F^a_{0i}-\epsilon_{ijk}B^j_aN^k\bigr)\nonumber\\
+\beta{N}(\hbox{det}B)^{1/2}\sqrt{\hbox{det}\Psi}\bigl(\Lambda+\hbox{tr}\Psi^{-1}\bigr)\biggr].
\end{eqnarray}
\noindent
Using (\ref{CANON9}) and the second equation of (\ref{CURVE1}), the equation of motion for $\Psi_{ae}$ is given by
\begin{eqnarray}
\label{CANON10}
{{\delta{I}_{Inst}} \over {\delta\Psi_{ae}}}
=B^i_e\bigl(F^a_{0i}-\epsilon_{ijk}B^j_aN^k\bigr)
-\beta{N}(\hbox{det}B)^{1/2}\sqrt{\hbox{det}\Psi}(\Psi^{-1}\Psi^{-1})^{ea}=0.
\end{eqnarray}
\noindent
Left multiplication of (\ref{CANON10}) by $B^{-1}$ gives the following equation
\begin{eqnarray}
\label{CURVE3}
F^a_{0i}-\bigl(\epsilon_{ijk}N^k+\beta\underline{N}(\hbox{det}B)(\hbox{det}\Psi)(\Psi^{-1}\Psi^{-1})^{ed}(B^{-1})^e_i(B^{-1})^d_j\bigr)B^j_a=0,
\end{eqnarray} 
\noindent 
where we have defined
\begin{eqnarray}
\label{CURVE4}
\underline{N}=N(\hbox{det}B)^{-1/2}(\hbox{det}\Psi)^{-1/2}.
\end{eqnarray}
\noindent 
We will ultimately regard equation (\ref{CURVE3}) as an evolution equation for the spatial connection $A^a_i$.  Equation (\ref{CURVE3}) also happens to be the 3+1 form of the following statement (See e.g. Appendix A for the details)
\begin{eqnarray} 
\label{CURVE5}
H_{(\beta)}^{\mu\nu\rho\sigma}F^a_{\rho\sigma}=0,
\end{eqnarray} 
\noindent 
namely that the field strength $F^a_{\mu\nu}$ is Hodge self-dual with respect to a spacetime metric $g_{\mu\nu}$ of signature $\beta$, having lapse function $N$, shift vector $N^i$, and spatial 3-metric $h_{ij}$ given by
\begin{eqnarray} 
\label{CURVE6}
h_{ij}=(\hbox{det}\Psi)(\Psi^{-1}\Psi^{-1})^{ae}(B^{-1})^a_i(B^{-1})^e_j(\hbox{det}B)
\end{eqnarray} 
\noindent 
where we have defined the Hodge duality operator
\begin{eqnarray} 
\label{CURVE7}
H^{\mu\nu\rho\sigma}_{(\beta)}
={1 \over 2}\sqrt{-g}\bigl(g^{\mu\rho}g^{\nu\sigma}-g^{\mu\sigma}g^{\nu\rho}\bigr)+{\beta \over 4}\epsilon^{\mu\nu\rho\sigma}.
\end{eqnarray}
\noindent 
The covariant derivative of (\ref{CURVE5}) is $D_{\nu}(H^{\mu\nu\rho\sigma}F^a_{\rho\sigma})=0$, which is the same as
\begin{eqnarray} 
\label{CURVE8}
D_{\nu}(\sqrt{-g}g^{\mu\rho}g^{\nu\sigma}F^a_{\rho\sigma})=-{\beta \over 2}\epsilon^{\mu\nu\rho\sigma}D_{\nu}F^a_{\rho\sigma}=0.
\end{eqnarray}
\noindent 
The second equality of (\ref{CURVE8}) is just the Bianchi identity, and the first equality is the Yang--Mills equation of motion for a $SU(2)$ Yang--Mills field $A^a_{\mu}$ coupled to some spacetime metric $g_{\mu\nu}$.  The fact that this Yang--Mills equation holds on account of the Bianchi identity suggests that the gauge connection $A^a_{\mu}$ describes the solution for a Yang--Mills instanton propagating on some spacetime with metric $g_{\mu\nu}$.\par 
\indent  
Subject to the definition (\ref{CURVE6}), equation (\ref{CURVE3}) can be written as
\begin{eqnarray} 
\label{STRESS5}
F^a_{0i}=\bigl(\epsilon_{ijk}N^k+\beta\underline{N}h_{ij}\bigr)B^j_a
\end{eqnarray} 
\noindent 
with $\underline{N}=Nh^{-1/2}$ with $h=\hbox{det}(h_{ij})$.  Right-multiplying (\ref{STRESS5}) with $B^{-1}$ and taking symmetric and antisymmetric parts, we have
\begin{eqnarray} 
\label{CURVE10}
N^i={1 \over 2}\epsilon^{ijk}F^a_{0j}(B^{-1})^a_k;~~\beta\underline{N}h_{ij}=F^a_{0(i}(B^{-1})^a_{j)}.
\end{eqnarray} 
\noindent 
So apparently, knowledge of the connection $A^a_{\mu}$ on a solution is sufficient to define a metric $g_{\mu\nu}$, up to a lapse function $N$.\footnote{Note that the signature of the spacetime metric $g_{\mu\nu}$ is also encoded in $N$, and a real $g_{\mu\nu}$ implies the following for $A^a_{\mu}$ in general.  For Euclidean signature all components components $A^a_{\mu}$ are allowed to be real-valued, whereas for Lorentzian signature at least some components must be complex in order to identically satisfy (\ref{STRESS5}).}
\par 
\indent 
To extract some more physical content from the equations, let us return to the level of (\ref{CANON10}).  Note that the antisymmetric part of (\ref{CANON10}) follows from contraction with $\epsilon_{dae}$ which gives
\begin{eqnarray}
\label{CANON14}
\epsilon_{dea}B^i_eF^a_{0i}=\epsilon_{ijk}\epsilon_{dea}B^i_eB^j_aN^k=2(\hbox{det}B)N^k(B^{-1})^d_k,
\end{eqnarray}
where we have used $\Psi_{ae}=\Psi_{(ae)}$ from the first equation of (\ref{CURVE1}).  This enables us to solve for the auxiliary field $N^i$ as
\begin{eqnarray}
\label{CANON15}
N^k={1 \over 2}\epsilon^{kij}F^a_{0i}(B^{-1})^a_j,
\end{eqnarray}
\noindent 
which is consistent with (\ref{CURVE10}).  For the symmetric part of the equation of motion (\ref{CANON10}), note upon 
defining $\epsilon^{0ijk}\equiv\epsilon^{ijk}$ and using the relation 
\begin{eqnarray}
\label{CANON12}
B^i_{(e}F^{a)}_{0i}={1 \over 2}\epsilon^{ijk}F^{(e}_{jk}F^{a)}_{0i}
={1 \over 8}F^a_{\mu\nu}F^e_{\rho\sigma}\epsilon^{\mu\nu\rho\sigma},
\end{eqnarray}
\noindent
that this is given by
\begin{eqnarray}
\label{CANON13}
{1 \over 8}F^a_{\mu\nu}F^e_{\rho\sigma}\epsilon^{\mu\nu\rho\sigma}+\beta{N}\sqrt{-g}(\Psi^{-1}\Psi^{-1})^{(ea)}=0,
\end{eqnarray}
\noindent
where $\sqrt{-g}=N\sqrt{h}=N\sqrt{\hbox{det}B}\sqrt{\hbox{det}\Psi}$ follows from taking the determinant of (\ref{CURVE6}).  Left and right multiplying (\ref{CANON13}) with $\Psi$, which is on-shell symmetric on account of the first equation of (\ref{CURVE1}), we obtain
\begin{eqnarray}
\label{OPTION762}
{1 \over 4}(\Psi^{bb^{\prime}}F^{b^{\prime}}_{\mu\nu})(\Psi^{ff^{\prime}}F^{f^{\prime}}_{\rho\sigma})\epsilon^{\mu\nu\rho\sigma}=-2\beta\sqrt{-g}\delta^{bf}.
\end{eqnarray}
\noindent
If one defines a two form $\Sigma^a={1 \over 2}\Sigma^a_{\mu\nu}{dx^{\mu}}\wedge{dx^{\nu}}$ with components $\Sigma^a_{\mu\nu}=\Sigma^a_{\mu\nu}[\Psi,A]$ according to the prescription
\begin{eqnarray}
\label{RECOGNIZES}
\Sigma^a_{\mu\nu}=\Psi_{ae}F^e_{\mu\nu},
\end{eqnarray}
\noindent
then equation (\ref{CANON13}) is equivalent to 
\begin{eqnarray}
\label{REDEF}
{1 \over 4}\Sigma^b_{\mu\nu}\Sigma^f_{\rho\sigma}\epsilon^{\mu\nu\rho\sigma}=-2\beta\sqrt{-g}\delta^{bf}.
\end{eqnarray}
\noindent
One recognizes (\ref{REDEF}) as none other than the simplicity constraint \cite{SELFDUAL}, namely the condition that the two forms $\Sigma^a$ can be constructed from wedge products of tetrad one forms $e^a=e^a_{\mu}dx^{\mu}$, with
\begin{eqnarray}
\label{RECOGNIZESES}
\Sigma^a=\Psi_{ae}F^e=\beta{e^0}\wedge{e^a}-{1 \over 2}\epsilon_{afg}{e^f}\wedge{e^g}\equiv({P}_{(\beta)})^a_{fg}{e^f}\wedge{e^g}.
\end{eqnarray}
\noindent
We have defined $(P_{(\beta)})^a_{fg}$ as a projection operator onto the self-dual combination of one-form wedge products for signature $\beta$, self-dual in the $SO(3)$ sense.  The trace of (\ref{REDEF}) fixes the volume form $\sqrt{-g}$ as
\begin{eqnarray} 
\label{THEVOLUME}
\sqrt{-g}=-{1 \over {24\beta}}\Sigma^b_{\mu\nu}\Sigma^b_{\rho\sigma}\epsilon^{\mu\nu\rho\sigma},
\end{eqnarray}
\noindent 
which must be nonzero.
\par

\section{Lagrange's equation for $A^a_{\mu}$}

Having extracted the physical content from the $\Psi$ equation of motion, we proceed next to the $A^a_{\mu}$ equation.  Toward that end it will be convenient to write the action (\ref{CANON9}) in a more covariant-looking from by separating $\Psi_{ae}$ into its symmetric and antisymmetric parts 
\begin{eqnarray} 
\label{SEPARATE}
\Psi_{ae}=\Psi_{(ae)}+{1 \over 2}\epsilon_{aed}\psi_d,
\end{eqnarray}
\noindent 
with $\psi_d$ given by (\ref{CURVE2}).  Note that the integrand of the $N^k$ term in (\ref{CANON9}) can be written as $\epsilon_{ijk}N^iB^j_eB^k_a\Psi_{ae}=(\hbox{det}B)N^i(B^{-1})^d_i\psi_d$.  Hence the action (\ref{CANON9}) can be written as
\begin{eqnarray}
\label{CANON17}
I_{Inst}=\int_Md^4x\biggl[{1 \over 8}\Psi_{ae}F^a_{\mu\nu}F^e_{\rho\sigma}\epsilon^{\mu\nu\rho\sigma}\nonumber\\
+\bigl({1 \over 2}\epsilon_{dae}F^a_{0i}B^i_e+N^i(B^{-1})^d_i\bigr)\psi_d
+\beta{N}(\hbox{det}B)^{1/2}\sqrt{\hbox{det}\Psi}\bigl(\Lambda+\hbox{tr}\Psi^{-1}\bigr)\biggr].
\end{eqnarray}
\noindent
The equation of motion for $N^i$ implies that $\psi_d=0$.  But since $\psi_d$ is also an independent dynamical field, then it is correct to set $\psi_d=0$ only after, and not before, writing down its Lagrange equation of motion 
\begin{eqnarray}
\label{CANON18}
{{\delta{I}_{Inst}} \over {\delta\psi_d}}\biggl\vert_{\psi_d=0}={1 \over 2}\epsilon_{dae}F^a_{0i}B^i_e+N^i(B^{-1})^d_i(\hbox{det}B)=0.
\end{eqnarray}
\noindent
Similarly, the equation of motion for $N$ is equivalent to $\Lambda+\hbox{tr}\Psi^{-1}=0$.  The solution to (\ref{CANON18}) is given precisely by (\ref{CANON15}).  The result is that the antisymmetric part of the equation of motion for $\Psi_{ae}$ is the same as the equation of motion for the antisymmetric part of $\Psi_{ae}$.\par 
\indent
To find the equation of motion for the connection $A^a_{\mu}$ it will be convenient to use the following relation $\epsilon_{ijk}N^iB^j_eB^k_a\Psi_{[ae]}={1 \over 2}\epsilon_{ijk}N^iB^j_aB^k_e\epsilon_{aed}\psi_d$.  Then the action (\ref{CANON17}) can also be written as
\begin{eqnarray}
\label{CANON19}
I_{Inst}=\int_Md^4x\biggl[{1 \over 8}\Psi_{ae}F^a_{\mu\nu}F^e_{\rho\sigma}\epsilon^{\mu\nu\rho\sigma}
+{1 \over 2}\epsilon_{dae}\bigl(F^a_{0j}B^j_e-\epsilon_{ijk}N^iB^j_eB^k_a\bigr)\psi_d\nonumber\\
-N(\hbox{det}B)^{1/2}\sqrt{\hbox{det}\Psi}\bigl(\Lambda+\hbox{tr}\Psi^{-1}\bigr)\biggr].
\end{eqnarray}
\noindent
The equation of motion for $A^a_{\mu}$ can be found by integration by parts of all terms containing the connection, which yields
\begin{eqnarray}
\label{CANON20}
{{\delta{I}} \over {\delta{A}^a_{\mu}}}=-\epsilon^{\mu\nu\rho\sigma}D_{\nu}(\Psi_{ae}F^e_{\rho\sigma})
-{1 \over 2}\delta^{\mu}_i\epsilon^{jml}D_m\biggl[\epsilon_{dag}\bigl(F^a_{0j}-2\epsilon_{jki}B^k_aN^i\bigr)\psi_d\nonumber\\
+(B^{-1})^g_jN(\hbox{det}B)^{1/2}\sqrt{\hbox{det}\Psi}\bigl(\Lambda+\hbox{tr}\Psi^{-1}\bigr)\biggr]=0.
\end{eqnarray}
\noindent
Note, due to the first and second equations of (\ref{CURVE1}), that the terms in large brackets in (\ref{CANON20}) vanish.  Therefore on-shell, (\ref{CANON20}) reduces to 
\begin{eqnarray}
\label{CANON21}
{{\delta{I}} \over {\delta{A}^a_{\mu}}}=\epsilon^{\mu\rho\sigma\nu}F^e_{\rho\sigma}D_{\nu}\Psi_{(ae)}=0,
\end{eqnarray}
\noindent
where we have used the Bianchi identity $\epsilon^{\mu\nu\rho\sigma}D_{\nu}F^e_{\rho\sigma}=0$.   

\subsection{Physical content of the equations} 

We will now extract the physical content of (\ref{CANON21}), which can be written in the following 3+1 form 
\begin{eqnarray} 
\label{CONTENT}
B^i_eD_i\Psi_{ae}=0;~~D_0\Psi_{ae}+\epsilon^{mij}(B^{-1})^e_mF^d_{0i}D_j\Psi_{ad}=0.
\end{eqnarray} 
\noindent 
The first equation of (\ref{CONTENT}) is the temporal part of (\ref{CANON21}), which is also the third constraint equation of (\ref{CURVE1}).  The second equation of of (\ref{CONTENT}) is $B^{-1}$ times the spatial part of (\ref{CANON21}),\footnote{This will be shown in a few lines as the quantity in medium-sized round brackets acted on by $D_i$ in the middle term of the equality on the right hand side of equation (\ref{RECOOGNIZES3}).} which appears to be an evolution equation for $\Psi_{ae}$.\footnote{The presence of the $F^d_{0i}$ term contaminates this with $A^a_i$ evolution parts.  We will return to this point later.}  Note, on account of the Bianchi identity, that the field strength can be brought inside the covariant derivative in (\ref{CANON21}), producing the equation 
\begin{eqnarray} 
\label{CONTENT1} 
\epsilon^{\mu\nu\rho\sigma}D_{\nu}(\Psi_{ae}F^e_{\rho\sigma})=0.
\end{eqnarray}
\noindent 
Acting on (\ref{CONTENT1}) with a covariant derivative $D_{\mu}$, we obtain
\begin{eqnarray} 
\label{CONTENT2}
\epsilon^{\mu\nu\rho\sigma}D_{\mu}D_{\nu}(\Psi_{ae}F^e_{\rho\sigma})
=f^{abc}\Psi_{ce}^bF^b_{\mu\nu}F^e_{\rho\sigma}\epsilon^{\mu\nu\rho\sigma}=0,
\end{eqnarray}
\noindent 
which invokes the definition of curvature as the commutator of two covariant derivatives.  To make progress, let us invoke the definition (\ref{RECOGNIZES}) in the following form
\begin{eqnarray}
\label{RECOOGNIZES}
F^a_{\mu\nu}=(\Psi^{-1})^{ae}\Sigma^e_{\mu\nu}.
\end{eqnarray}
\noindent 
Then substitution of (\ref{RECOOGNIZES}) into (\ref{CONTENT2}) yields
\begin{eqnarray} 
\label{RECOOGNIZES1}
f^{abc}(\Psi^{-1})^{bb^{\prime}}\Sigma^{b^{\prime}}_{\mu\nu}\Sigma^c_{\rho\sigma}\epsilon^{\mu\nu\rho\sigma}
=-8\beta\sqrt{-g}f^{abc}(\Psi^{-1})^{bc}=0.
\end{eqnarray} 
\noindent 
The first equality of (\ref{RECOOGNIZES1}) follows from (\ref{REDEF}) and the second equality follows from $\psi_d=0$, the first equation of (\ref{CURVE1}).  So we see that the equations of motion of $A^a_{\mu}$ and those of $\Psi_{ae}$ are mutually consistent.\par 
\indent 
Let us now analyse the meaning of (\ref{CONTENT2}) in 3+1 form, using the equivalent version obtainable from (\ref{CANON21})
\begin{eqnarray} 
\label{RECOOGNIZES2}
D_{\mu}(\epsilon^{\mu\nu\rho\sigma}F^e_{\rho\sigma}D_{\sigma}\Psi_{ae})
=D_0(\epsilon^{0ijk}F^e_{ij}D_k\Psi_{ae})\nonumber\\
+D_i(\epsilon^{i0jk}F^e_{0j}D_k\Psi_{ae})
+D_i(\epsilon^{ij0k}F^e_{j0}D_k\Psi_{ae})+D_i(\epsilon^{ijk0}F^e_{jk}D_0\Psi_{ae})=0.
\end{eqnarray}
\noindent 
Using the definitions (\ref{CURVE}), then this is the same thing as
\begin{eqnarray} 
\label{RECOOGNIZES3}
{1 \over 2}D_{\mu}(\epsilon^{\mu\nu\rho\sigma}F^e_{\rho\sigma}D_{\sigma}\Psi_{ae})
=D_0(B^k_eD_k\Psi_{ae})-D_i\bigl(\epsilon^{ijk}F^d_{0j}D_k\Psi_{ad}+B^i_eD_0\Psi_{ae}\bigr)=0.
\end{eqnarray}
\noindent
Using the definition $G_a=B^i_eD_i\Psi_{ae}$ from the third equation of (\ref{CURVE1}) and the definition of the covariant derivative in (\ref{COVARIANTDERIVATIVE}), then equation (\ref{RECOOGNIZES3}) can be written as 
\begin{eqnarray} 
\label{RECOOGNIZES4}
\dot{G}_a=-f_{abc}A^b_0G_c+D_i\bigl(\epsilon^{ijk}F^d_{0j}D_k\Psi_{ad}+B^i_eD_0\Psi_{ae}\bigr)=0.
\end{eqnarray}
\noindent 
Recall that $G_a=0$ which is the Gauss' law constraint, the third equation of (\ref{CURVE1}) which is the temporal part of the Lagrange's equation for $A^a_{\mu}$, and that the quantity in round brackets in (\ref{RECOOGNIZES4}) is just the spatial part of the equation for $A^a_{\mu}$.  Equation (\ref{RECOOGNIZES4}) states simply that if $G_a=0$ holds at some initial time $t_0$, then as long as the spatial $A^a_i$ equation of motion holds, then the time derivative $\dot{G}_a$ is also zero.  In other words, the Gauss' law constraint is preserved for all time by the evolution equations.\footnote{Hence, to have a solution for $I_{Inst}$, one essentially needs only to find a symmetric $\Psi$ with $\hbox{tr}\Psi^{-1}=-\Lambda$, which satisfies the Gauss' law constraint $G_a(0)=G_a[\Psi(0),A(0)]=0$, where we have suppressed the spatial dependence.  Then the evolution equations guarantee that the variables $\Psi(t),A(t)$ solve Gauss' law $G_a(t)=0$ at any time $t$.}\par 
\indent 
On a final note, observe that equation (\ref{CONTENT1}) can also be written as
\begin{eqnarray} 
\label{RECOOGNIZES5}
\epsilon^{\mu\nu\rho\sigma}D_{\nu}\Sigma^a_{\rho\sigma}=0,
\end{eqnarray}
\noindent
which uses (\ref{RECOGNIZES}).  We will examine the implications of this in the next section.

\section{Recapitulation}

We will now reflect on the results of this paper thus far, to put them into context.  Starting from the action
\begin{eqnarray} 
\label{STARTINGFROM}
I_{Inst}=\int{dt}\int_{\Sigma}d^3x\biggl[\Psi_{ae}B^i_e\dot{A}^a_i+A^a_0B^i_eD_i\Psi_{ae}\nonumber\\
-\epsilon_{ijk}N^iB^j_eB^k_a\Psi_{ae}+\beta\sqrt{\hbox{det}B}\sqrt{\hbox{det}\Psi}\bigl(\Lambda+\hbox{tr}\Psi^{-1}\bigr)\biggr],
\end{eqnarray}
\noindent 
we have shown that $A^a_{\mu}$ satisfies the Yang--Mills equations of motion in a curved spacetime $g_{\mu\nu}$, and that $\Psi_{ae}$ is a symmetric matrix with eigenvalues $\lambda_1,\lambda_2,\lambda_3$, which is constrained by the two conditions
\begin{eqnarray} 
\label{STARTINGFROM1}
\Lambda+{1 \over {\lambda_1}}+{1 \over {\lambda_2}}+{1 \over {\lambda_3}}=0;~~B^i_eD_i\Psi_{ae}=0.
\end{eqnarray} 
\noindent 
Additionally, the field strength $F^a_{\mu\nu}$ is Hodge self-dual with $g_{\mu\nu}=g_{\mu\nu}[A,\Psi]$ constructible completely from the connection $A^a_{\mu}$ and the field $\Psi_{ae}$.  But we have not commented on the significance of this metric $g_{\mu\nu}$.\par 
\indent 
Recall the re-definition of variables (\ref{RECOGNIZES}), which transforms the pertinent equations of motion into (\ref{REDEF}) and (\ref{RECOOGNIZES5}).  Using (\ref{THEVOLUME}) in conjunction with these equations, we have 
\begin{eqnarray} 
\label{STARTINGFROM2}
\Sigma^a_{\mu\nu}=\Psi_{ae}F^e_{\mu\nu};\nonumber\\
\Sigma^b_{\mu\nu}\Sigma^f_{\rho\sigma}\epsilon^{\mu\nu\rho\sigma}-{1 \over 3}\delta^{bf}(\Sigma^b_{\mu\nu}\Sigma^f_{\rho\sigma}\epsilon^{\mu\nu\rho\sigma})=0;\nonumber\\
\epsilon^{\mu\nu\rho\sigma}D_{\nu}\Sigma^a_{\rho\sigma}=0,
\end{eqnarray}  
\noindent 
re-written here for completeness.  Say that we define the two forms
\begin{eqnarray} 
\label{TWOSTEP}
\Sigma^a={1 \over 2}\Sigma^a_{\mu\nu}{dx^{\mu}}\wedge{dx^{\nu}};~~F^a={1 \over 2}F^a_{\mu\nu}{dx^{\mu}}\wedge{dx^{\nu}},
\end{eqnarray} 
\noindent
with $F^a$ being the curvature two form for the connection one form $A^a=A^a_{\mu}dx^{\mu}$.  Then equations (\ref{STARTINGFROM2}) are none other than the equations of motion from Plebanski's theory of gravity with action
\begin{eqnarray}
\label{PLEBANSKI}
I_{Pleb}=\int_M\delta_{ae}{\Sigma^a}\wedge{F^e}-{1 \over 2}(\delta_{ae}\varphi+\psi_{ae}){\Sigma^a}\wedge{\Sigma^e},
\end{eqnarray}
\noindent
where $\psi_{ae}$ is symmetric and traceless and $\varphi=-{\Lambda \over 3}$ is a numerical constant.  It is known, when Plebanski's equations of motion are satisfied, that Plebanski's equations imply the Einstein equations \cite{PLEBANSKI}.\par 
\indent  
This provides the rationale for referring to (\ref{STARTINGFROM}) as the instanton representation of Plebanski gravity.  It is precisely because the action (\ref{STARTINGFROM}) yields the same equations of motion as Plebanski's theory \cite{PLEBANSKI}, while describing Yang--Mills instantons.  Since the metric $g_{\mu\nu}$ which the Yang--Mills theory couples and is Hodge self-dual with respect to is constructible from the very same connection $A^a_{\mu}$ combined with $\Psi_{ae}$, then one has a representation of Plebanski's gravity which admits gravitational instantons as solutions.  So in a sense, when (\ref{NONDEGENERATE}) holds, (\ref{STARTINGFROM}) carries the interpretation of an alternate 3+1 formulation of Plebanksi's theory to the Ashtekar formalism.

\subsection{Evolution equations for the basic variables}

Now that we have demonstrated that $I_{Inst}$ is indeed a theory of gravity, we will now derive its complete evolution equations for the basic fields $A^a_i,\Psi_{ae}$.  We have thus far obtained (\ref{CURVE3}) and the second equation of (\ref{CONTENT}), repeated here for completeness
\begin{eqnarray}
\label{CANON24}
F^a_{0i}-\epsilon_{ijk}B^j_aN^k-\beta{N}(\hbox{det}B)^{1/2}\sqrt{\hbox{det}\Psi}(B^{-1})^e_i(\Psi^{-1}\Psi^{-1})^{ea}=0;\nonumber\\
D_0\Psi_{ae}=-\epsilon^{ijk}(B^{-1})^e_iF^g_{0j}D_k\Psi_{ag}.
\end{eqnarray}
\noindent
The first line of (\ref{CANON24}) is an evolution equation for $A^a_i$, whereas the second equation if not for the $F^g_{0j}$ part would be a separate evolution equation for $\Psi_{ae}$.  To rectify this we will substitute $F^g_{0j}$ from the first equation of (\ref{CANON24}) into the second equation, which yields
\begin{eqnarray}
\label{CANON25}
D_0\Psi_{ae}=-\epsilon^{ijk}(B^{-1})^e_i\epsilon_{jmn}B^m_gN^nD_k\Psi_{ag}\nonumber\\
-\beta{N}(\hbox{det}B)^{1/2}\sqrt{\hbox{det}\Psi}\epsilon^{ijk}(B^{-1})^e_i(B^{-1})^f_j(\Psi^{-1}\Psi^{-1})^{fg}D_k\Psi_{ag}.
\end{eqnarray}
\noindent
Applying epsilon identities and the definition of determinants to (\ref{CANON25}), we have
\begin{eqnarray}
\label{CANON26}
D_0\Psi_{ae}=-\bigl(\delta^k_m\delta^i_n-\delta^k_n\delta^i_m\bigr)(B^{-1})^e_iB^m_gN^nD_k\Psi_{ag}\nonumber\\
-\beta{N}(\hbox{det}B)^{-1/2}\epsilon^{efd}(\Psi^{-1}\Psi^{-1})^{fg}B^k_dD_k\Psi_{ag}\nonumber\\
=-N^i(B^{-1})^e_iB^k_gD_k\Psi_{ag}+N^kD_k\Psi_{ae}
-\beta{N}(\hbox{det}B)^{-1/2}\epsilon^{efd}(\Psi^{-1}\Psi^{-1})^{fg}B^k_dD_k\Psi_{ag}.
\end{eqnarray}
Note that the first term on the right hand side of (\ref{CANON26}) is directly proportional to the Gauss' function $G_a$, which vanishes when $G_a=0$.  Using the relation from (\ref{COVARIANTDERIVATIVE})
\begin{eqnarray}
\label{THERELATION}
D_0\Psi_{ae}=\dot{\Psi}_{ae}+A^b_0\bigl(f_{abc}\Psi_{ce}+f_{ebc}\Psi_{ac}\bigr), 
\end{eqnarray}
\noindent
then we can separate the part of (\ref{CANON26}) containing $\dot{\Psi}_{ae}$ from the $A^b_0$ part.  The result is that the $I_{Inst}$ equations of motion imply the evolution equations
\begin{eqnarray}
\label{CANON28}
\dot{A}^a_i=D_iA^a_0+\epsilon_{ijk}B^j_aN^k+\beta{N}(\hbox{det}B)^{1/2}\sqrt{\hbox{det}\Psi}(B^{-1})^b_i(\Psi^{-1}\Psi^{-1})^{ba};\nonumber\\
\dot{\Psi}_{ae}=-A^b_0\bigl(f_{abc}\Psi_{ce}+f_{ebc}\Psi_{ac}\bigr)+N^kD_k\Psi_{ae}\nonumber\\
-\beta{N}(\hbox{det}B)^{1/2}\sqrt{\hbox{det}\Psi}\bigl[(\hbox{det}B)^{-1}\epsilon^{efd}(\Psi^{-1}\Psi^{-1})^{fb}B^k_dD_k\Psi_{ab}\bigr].
\end{eqnarray}
\noindent
Note that the second equation of (\ref{CANON28}) is valid if and only if $\Psi_{[ae]}=0$, the consistency of which can be checked by examining its antisymmetric part.  Contracting this equation 
with $\epsilon_{gae}$ yields $\epsilon_{gae}\dot{\Psi}_{ae}$ for the left hand side.  The right hand side splits into two terms which we will in turn analyse.  The term $N^k\partial_k(\epsilon_{gae}\Psi_{ae})$ in the covariant derivative is automatically zero when $\Psi_{[ae]}$ is zero.  The antisymmetric part of all the $A^b_0,N^i$ terms is of the form 
\begin{eqnarray}
\label{ANTISYM}
\epsilon_{dae}\bigl(f_{abc}\Psi_{ce}+f_{ebc}\Psi_{ac}\bigr)(A^b_0-N^kA^b_k)\nonumber\\
=\bigl(\bigl(\delta_{eb}\delta_{dc}-\delta_{ec}\delta_{db}\bigr)\Psi_{ce}+\bigl(\delta_{bd}\delta_{ca}-\delta_{ba}\delta_{cd}\bigr)\Psi_{ac}\bigr)(A^b_0-N^kA^b_k)\nonumber\\
=2\Psi_{[bd]}(A^b_0-N^kA^b_k),
\end{eqnarray}
\noindent
which is also proportional to $\Psi_{[ae]}$.  This leaves remaining the term involving $N$, whose antisymmetric part up to multiplicative factors is
\begin{eqnarray}
\label{ANTISYM1}
\epsilon_{gae}\epsilon^{efd}(\Psi^{-1}\Psi^{-1})^{fc}B^k_dD_k\Psi_{ab}
=\bigl(\delta^f_g\delta^d_a-\delta^f_a\delta^d_g\bigr)(\Psi^{-1}\Psi^{-1})^{fb}B^k_dD_k\Psi_{ab}\nonumber\\
=(\Psi^{-1}\Psi^{-1})^{gb}B^k_aD_k\Psi_{ab}-(\Psi^{-1}\Psi^{-1})^{fb}B^k_gD_k\Psi_{fb}.
\end{eqnarray}
\noindent
Note that the first term on the right hand side of (\ref{ANTISYM1}) is directly proportional to the Gauss' constraint $G_a$.  The second term can be written as
\begin{eqnarray}
\label{ANTISYM2}
(\Psi^{-1}\Psi^{-1})^{fb}B^k_gD_k\Psi_{fb}
=B^k_d\partial_k(\Lambda+\hbox{tr}\Psi^{-1})
\end{eqnarray}
\noindent  
when $\Psi_{ae}$ is symmetric.  Note that we have appended $\Lambda$ as a constant of spatial integration.  This permits the identification of the covariant derivative with the spatial derivative of the Hamiltonian constraint which is a gauge scalar.  This term also vanishes on-shell.  The result is that the previous manipulations involving $\Psi_{[ae]}$ are valid when (\ref{CURVE1}) holds.\par 
\indent 
We have shown that the first and third equations of (\ref{CURVE1}) are preserved by the evolution equations, provided that the second equation of (\ref{CURVE1}) holds.  All that remains is to show that the second equation of (\ref{CURVE1}) is also preserved under time evolution.  To accomplish this it will suffice to use the $\Psi_{ae}$ evolution equation in conjunction with the identity
\begin{eqnarray} 
\label{SUFFICE}
\partial_{\mu}\bigl(\Lambda+\hbox{tr}\Psi^{-1}\bigr)=\partial_{\mu}\hbox{tr}\Psi^{-1}=-(\Psi^{-1}\Psi^{-1})^{ea}\partial_{\mu}\Psi_{ae}.
\end{eqnarray} 
\noindent 
The $\Psi_{ae}$ evolution equation can be written as
\begin{eqnarray} 
\label{SUFFICE1}
\dot{\Psi}_{ae}-N^k\partial_k\Psi_{ae}
=(A^b_0-N^kA^b_k)\bigl(f_{abc}\Psi_{ce}+f_{ebc}\Psi_{ac}\bigr)\nonumber\\
-\beta{N}\sqrt{{{\hbox{det}\Psi} \over {\hbox{det}B}}}
\epsilon^{efd}(\Psi^{-1}\Psi^{-1})^{fb}B^k_dD_k\Psi_{ab}.
\end{eqnarray}
\noindent 
Contraction of (\ref{SUFFICE1}) by $(\Psi^{-1}\Psi^{-1})^{ea}$ for the left hand side yields the time derivative of the second equation of (\ref{CURVE1}) and its spatial derivative, the latter of which vanishes when the constraint holds.  For this constraint to be preserved under time evolution, then the right hand side of (\ref{SUFFICE1}) must be zero under this contraction.  It suffices to analyze the pertinent parts in turn.\par 
\indent 
For the first term on the right hand side of (\ref{SUFFICE1}) we have
\begin{eqnarray} 
\label{SUFFICE2}
(\Psi^{-1}\Psi^{-1})^{ea}\bigl(f_{abc}\Psi_{ce}+f_{ebc}\Psi_{ac}\bigr)=f_{abc}(\Psi^{-1})^{ca}+f_{ebc}(\Psi^{-1})^{ec}=0
\end{eqnarray}
\noindent 
on account of $\psi_d=0$.  For the second term, using the definition of determinants we have
\begin{eqnarray} 
\label{SUFFICE3}
\epsilon^{efd}(\Psi^{-1}\Psi^{-1})^{ea}(\Psi^{-1}\Psi^{-1})^{fb}B^k_dD_k\Psi_{ab}
=(\hbox{det}\Psi)^{-2}(\Psi\Psi)^{de}B^k_dD_k(\epsilon^{abc}\Psi_{ab})=0,
\end{eqnarray}
\noindent 
again, which is valid for $\psi_d=0$.  The result is that the initial value constraints (\ref{CURVE1}) are preserved under time evolution by the evolution equations.

\section{Summary and future research}

We have introduced a formulation for gravity where the basic fields are a $SO(3)$ gauge connection and a 3 by 3 matrix, $A^b_{\mu}$ and $\Psi_{ae}$ respectively.  We have named this theory "the instanton representation of Plebanski gravity" with action $I_{Inst}$, written in 3+1 form as
\begin{eqnarray} 
\label{BOOTINST}
I_{Inst}=\int{dt}\int_{\Sigma}d^3x\biggl[\Psi_{ae}B^i_e\dot{A}^a_i+A^a_0B^i_eD_i\Psi_{ae}\nonumber\\
-\epsilon_{ijk}N^iB^j_eB^k_a\Psi_{ae}+\beta\sqrt{\hbox{det}B}\sqrt{\hbox{det}\Psi}\bigl(\Lambda+\hbox{tr}\Psi^{-1}\bigr)\biggr].
\end{eqnarray}
\noindent
As an aside, note that it is possible to eliminate $\psi_d$ and $N^i$ by evaluating the action (\ref{BOOTINST}) on the critical point $\psi_d=0,N^i={1 \over 2}\epsilon^{ijk}F^a_{0j}(B^{-1})^a_k$, which leads to the reduced action 
\begin{eqnarray}
\label{CANON22}
I_0=I_{Inst}\biggl\vert_{\psi_d=0,N^i={1 \over 2}\epsilon^{ijk}F^a_{0j}(B^{-1})^a_k}\nonumber\\
=\int_Md^4x\biggl[{1 \over 8}\Psi_{ae}F^a_{\mu\nu}F^e_{\rho\sigma}\epsilon^{\mu\nu\rho\sigma}+\eta\bigl(\Lambda+\hbox{tr}\Psi^{-1}\bigr)\biggr]
\end{eqnarray}
\noindent
where $\eta=\sqrt{\hbox{det}B}\sqrt{\hbox{det}\Psi}$, whence $\Psi_{ae}=\Psi_{(ae)}$ is now symmetric.  Note that the $\Psi_{ae}$ equation of motion of (\ref{CANON22}) is precisely the symmetric part of the equation of motion for $\Psi_{ae}$ from 
(\ref{BOOTINST}), and moreover that the $A^a_{\mu}$ equation of motion from (\ref{CANON21}) and (\ref{BOOTINST}) are the same when $\psi_d=0$.\par 
\indent 
From (\ref{CANON22}) one can eliminate $\Psi_{(ae)}$ through its equation of motion and substitute back into (\ref{CANON22}), obtaining the pure spin connection formulation of gravity due to CDJ \cite{SPINCON}.  But what is missing at the level of (\ref{CANON22}) in relation to (\ref{BOOTINST}), is that the latter enables a determination of the shift vector $N^i$ and also produces the Hodge duality condition, whereas the former does not.\footnote{This shows that (\ref{BOOTINST}), which is sometimes erroneously referred to as part of the "CDJ formalism" on the basis of its resemblance to (\ref{CANON22}), is indeed not the same formalism.  Hence we distinguish (\ref{BOOTINST}) through the term, "instanton representation of Plebanski gravity".}\par 
\indent
We have chosen the term "instanton representation of Plebanski gravity" for (\ref{BOOTINST}) because the Plebanski equations of motion can be recovered from $I_{Inst}$, which also admits Yang--Mills instantons (gravitational instantons in consequence) as solutions.  The equations of motion for $I_{Inst}$ in 3+1 form are given by the constraint equations
\begin{eqnarray} 
\label{FINALLY}
B^i_eD_i\Psi_{ae}=0;~~\epsilon_{dae}\Psi_{ae}=0;~~\Lambda+\hbox{tr}\Psi^{-1}=0,
\end{eqnarray} 
\noindent 
combined with the evolution equations
\begin{eqnarray}
\label{FINALLY1}
\dot{A}^a_i=D_iA^a_0+\epsilon_{ijk}B^j_aN^k+\beta{N}(\hbox{det}B)^{1/2}\sqrt{\hbox{det}\Psi}(B^{-1})^b_i(\Psi^{-1}\Psi^{-1})^{ba};\nonumber\\
\dot{\Psi}_{ae}=-A^b_0\bigl(f_{abc}\Psi_{ce}+f_{ebc}\Psi_{ac}\bigr)+N^kD_k\Psi_{ae}\nonumber\\
-\beta{N}(\hbox{det}B)^{1/2}\sqrt{\hbox{det}\Psi}\bigl[(\hbox{det}B)^{-1}\epsilon^{efd}(\Psi^{-1}\Psi^{-1})^{fb}B^k_dD_k\Psi_{ab}\bigr].
\end{eqnarray}
\noindent 
Equations (\ref{FINALLY1}) follow from the covariant form of the equations
\begin{eqnarray} 
\label{FINALLY2}
H^{\mu\nu\rho\sigma}_{(\beta)}F^a_{\rho\sigma}=0;~~\epsilon^{\mu\nu\rho\sigma}F^e_{\nu\rho}D_{\sigma}\Psi_{ae}=0,
\end{eqnarray} 
\noindent 
namely a Hodge duality condition with respect to a spacetime metric $g_{\mu\nu}=g_{\mu\nu}[A,\Psi;N]$, and a derivative condition on $\Psi_{ae}$.  The constraints (\ref{FINALLY}) are preserved under time evolution by the evolution equations (\ref{FINALLY1}), (\ref{FINALLY2}), signifying that $I_{Inst}$ is a self-consistent theory of gravity when $(\hbox{det}\Psi)$ and $(\hbox{det}B)$ are both nonzero.\par 
\indent 
We will now provide some of the motivations for introducing the instanton representation of Plebanski gravity. (i) Due to its satisfaction of the Yang--Mills equations as a consequence of Hodge duality, it is conducive to a systematic categorization and construction of gravitational instanton solutions.  Gravitational instantons are important in Euclidean quantum gravity, where their on-shell action constitutes a dominant contribution to the path integral. (ii) The initial value constraints (\ref{FINALLY}) are more tractable, since they consist of only three differential equations and four simple algebraic equations.  The solution for $I_{Inst}$ reduces essentially to the ability to solve the Gauss' constraint $G_a=0$, which we will relegate as an area of future research.  (iii) In the evolution equations (\ref{FINALLY1}) spatial derivatives act only on $A^a_0$, but not on $N,N^i$.  This would render gauge-fixing procedures more tractable, which in combination with (ii) suggests that (\ref{BOOTINST}) has the potential to make accessible larger sectors of the reduced phase space for gravity in relation to other approaches.  These properties are particularly useful for numerical relativity and in general for constructing new General Relativity solutions which existing approaches might miss.\par 
\indent 
On a final note, on page 1090 of \cite{PELDAN} there appears a diagram depicting a library of various actions for gravity.  In the top left-hand corner, the only action depicted from which Plebanski's action can be derived is the self-dual Hilbert--Palatini action.  The results of the present paper show that \cite{PELDAN} can be updated by inclusion of a second two-way arrow from the Plebanski's action box to $I_{Inst}$, given by (\ref{BOOTINST}).  The utilization of $I_{Inst}$ as a practical tool in the construction and in the interpretation of General Relativity solutions will constitute an area for on-going and for future research.

\section{Appendix A: Verification of the Hodge duality condition in 3+1 form}

We will now show that equation (\ref{STRESS5}), re-written here for completeness
\begin{eqnarray} 
\label{COMPLETENESSES}
F^a_{0i}+\bigl(\epsilon_{ijk}N^k+\beta\underline{N}h_{ij}\bigr)B^j_a=0,
\end{eqnarray} 
\noindent
is indeed the 3+1 form of the statement of Hodge duality of the field strength $F^a_{\mu\nu}$ with respect to the 
metric $g_{\mu\nu}$, whose covariant form in 3+1 form is 
\begin{eqnarray}
\label{SELFISH}
g^{00}=-{1 \over {N^2}};~~g^{0i}={{N^i} \over {N^2}};~~g^{ij}=h^{ij}-{{N^iN^j} \over {N^2}}.
\end{eqnarray}
\noindent
To show this, we will derive the Hodge self-duality condition for Yang--Mills theory in curved spacetime, using the 3+1 decomposition (\ref{SELFISH}).  The Hodge self-duality condition for $F^a_{\mu\nu}$ can be written in the form
\begin{eqnarray}
\label{SELFISH1}
\sqrt{-g}g^{\mu\rho}g^{\nu\sigma}F^a_{\rho\sigma}=-{\beta \over 2}\epsilon^{\mu\nu\rho\sigma}F^a_{\rho\sigma},
\end{eqnarray}
\noindent
where $\beta$ is a numerical constant to be fixed by a consistency condition.  It will also be convenient to use $B^i_a={1 \over 2}\epsilon^{ijk}F^a_{jk}$ for the spatial part of the field strength $F^a_{\mu\nu}$.  Expanding (\ref{SELFISH1}) and using $F^a_{00}=0$, we have
\begin{eqnarray}
\label{SELFISH2}
N\sqrt{h}\bigl(\bigl(g^{\mu0}g^{\nu{j}}-g^{\nu0}g^{\mu{j}}\bigr)F^a_{0j}+g^{\mu{i}}g^{\nu{j}}\epsilon_{ijk}B^k_a\bigr)
=-{\beta \over 2}\bigl(2\epsilon^{\mu\nu{0}i}F^a_{0i}+\epsilon^{\mu\nu{ij}}\epsilon_{ijm}B^m_a\bigr).
\end{eqnarray}
\noindent
We will now examine the components of (\ref{SELFISH2}) in turn.  Note that the time-time component $\mu=0,\nu=0$ yields $0=0$, which is trivially satisfied.  So we may move on to components involving the spatial indices.

\subsection{Space-time components}

Moving on to the $\mu=0,\nu=k$ component of (\ref{SELFISH2}), we have
\begin{eqnarray}
\label{FROMTHE}
N\sqrt{h}\Bigl(\bigl(g^{00}g^{kj}-g^{k0}g^{0j}\bigr)F^a_{0j}+g^{0i}g^{kj}\epsilon_{ijm}B^m_a\Bigr)=-\beta{B}^k_a
\end{eqnarray}
\noindent
Inserting the metric components from (\ref{SELFISH}) into (\ref{FROMTHE}), we have
\begin{eqnarray}
\label{FROMTHE1}
N\sqrt{h}\Bigl(\Bigl(-{1 \over {N^2}}\Bigr)\Bigl(h^{kj}-{{N^kN^j} \over {N^2}}\Bigr)-\Bigl({{N^kN^j} \over {N^2}}\Bigr)\Bigr)F^a_{0j}\nonumber\\
+N\sqrt{h}\Bigl({{N^i} \over {N^2}}\Bigr)\Bigl(h^{kj}-{{N^kN^j} \over {N^2}}\Bigr)\epsilon_{ijm}B^m_a\Bigr)=-\beta{B}^k_a.
\end{eqnarray}
\noindent
Cancelling off the terms multipying $F^a_{0j}$ which are quadratic in $N^i$, we have
\begin{eqnarray}
\label{FROMTHE2}
-\Bigl({{\sqrt{h}} \over N}\Bigr)h^{kj}\bigl(F^a_{0j}-\epsilon_{jmi}B^m_aN^i\bigr)=-\beta{B}^k_a.
\end{eqnarray}
\noindent
Multiplying (\ref{FROMTHE2}) by $\underline{N}=N/\sqrt{h}$ and by $h_{lk}$, this yields
\begin{eqnarray}
\label{FROMTHE3}
F^a_{0l}-\epsilon_{lmi}B^m_aN^i-\beta\underline{N}h_{lk}B^k_a=0.
\end{eqnarray}
\noindent
Equation (\ref{FROMTHE3}) is the same as (\ref{STRESS5}), which confirms that we are on the right track.  Since we must verify Hodge duality on all components, then we must show that (\ref{FROMTHE3}) is consistent with the condition of Hodge duality with respect to the spatial components.

\subsection{Space-space components}

Moving on to the space-space component we substitute $\mu=m$ and $\nu=n$ in equation (\ref{SELFISH2}), which yields
\begin{eqnarray}
\label{FROMTHE4}
N\sqrt{h}\Bigl(\bigl(g^{m0}g^{nj}-g^{n0}g^{mj}\bigr)F^a_{0j}+g^{mi}g^{nj}\epsilon_{ijk}B^k_a\Bigr)=-\beta\epsilon^{mn0j}F^a_{0j}.
\end{eqnarray}
\noindent
Inserting the metric components from (\ref{SELFISH}) into (\ref{FROMTHE4}), we have
\begin{eqnarray}
\label{FROMTHE5}
N\sqrt{h}\Bigl[\Bigl({{N^m} \over {N^2}}\Bigr)\Bigl(h^{nj}-{{N^nN^j} \over {N^2}}\Bigr)-\Bigl({{N^n} \over {N^2}}\Bigr)\Bigl(h^{mj}-{{N^mN^j} \over {N^2}}\Bigr)\Bigr)F^a_{0j}\nonumber\\
+\Bigl(h^{mi}-{{N^mN^i} \over {N^2}}\Bigr)\Bigl(h^{nj}-{{N^nN^j} \over {N^2}}\Bigr)\epsilon_{ijk}B^k_a\Bigr]=-\beta\epsilon^{0mnj}F^a_{0j}.
\end{eqnarray}
\noindent
Equation (\ref{FROMTHE5}), upon vanishing of the term quartic in the shift vector $N^i$, yields
\begin{eqnarray}
\label{FROMTHE6}
{{\sqrt{h}} \over N}\bigl(h^{nj}N^m-h^{mj}N^n\bigr)F^a_{0j}+N\sqrt{h}h^{mi}h^{nj}\epsilon_{ijk}B^k_a\nonumber\\
-{{\sqrt{h}} \over N}\bigl(h^{mi}N^nN^j+h^{nj}N^mN^i\bigr)\epsilon_{ijk}B^k_a=-\beta\epsilon^{0mnj}F^a_{0j}.
\end{eqnarray}
\noindent
From the third term on the left hand side of (\ref{FROMTHE6}), we have the following relation upon relabelling indices $i\leftrightarrow{j}$ on the first term in brackets
\begin{eqnarray}
\label{FROMTHE7}
-h^{mi}N^nN^j\epsilon_{ijk}B^k_a-h^{nj}N^mN^i\epsilon_{ijk}B^k_a=-h^{mj}N^nN^i\epsilon_{jik}B^k_a-h^{nj}N^mN^i\epsilon_{ijk}B^k_a\nonumber\\
=\epsilon_{ijk}\bigl(h^{mj}N^n-h^{nj}N^m\bigr)N^iB^k_a.
\end{eqnarray}
\noindent
Note that the combination $h^{nj}N^m-h^{mj}N^n$ on the right hand side of (\ref{FROMTHE7}) is the same term multiplying $F^a_{0j}$ in the left hand side of (\ref{FROMTHE6}).  Using this fact, then (\ref{FROMTHE6}) can be written as
\begin{eqnarray}
\label{FROMTHE8}
{{\sqrt{h}} \over N}\Bigl[\bigl(h^{nj}N^m-h^{mj}N^n\bigr)\bigl(F^a_{0j}-\epsilon_{jki}B^k_aN^i\bigr)\Bigr]+\underline{N}\epsilon^{mnl}h_{lk}B^k_a=-\beta\epsilon^{mnj}F^a_{0j}
\end{eqnarray}
\noindent
where $\epsilon^{0mnj}=\epsilon^{mnj}$.  Using $F^a_{0j}-\epsilon_{jki}B^k_aN^i=\beta\underline{N}h_{jk}B^k_a$ from (\ref{FROMTHE3}) in (\ref{FROMTHE8}), then we have
\begin{eqnarray}
\label{FROMTHE9}
{{\sqrt{h}} \over N}\bigl(h^{mj}N^n-h^{nj}N^m\bigr)\beta\underline{N}h_{jk}B^k_a+\underline{N}\epsilon^{mnl}h_{lk}B^k_a=-\beta\epsilon^{mnj}F^a_{0j}.
\end{eqnarray}
\noindent
This simplifies to
\begin{eqnarray}
\label{FROMTHE10}
\beta\bigl(\delta^n_kN^m-\delta^m_kN^n\bigr)B^k_a+\underline{N}\epsilon^{mnl}h_{lk}B^k_a=-\beta\epsilon^{mnj}F^a_{0j}\nonumber\\
\longrightarrow-\beta\bigl(\epsilon^{mnj}F^a_{0j}-B^m_aN^n+B^n_aN^m\bigr)=\underline{N}\epsilon^{mnj}h_{jk}B^k_a.
\end{eqnarray}
\noindent
Contracting (\ref{FROMTHE10}) with $\epsilon_{mnl}$ and dividing by $2$, we obtain the relation
\begin{eqnarray}
\label{FROMTHE11}
F^a_{0l}-\epsilon_{lmn}B^m_aN^n+{1 \over \beta}\underline{N}h_{lk}B^k_a=0.
\end{eqnarray}
\noindent
To avoid a contradiction, Consistency of (\ref{FROMTHE11}) with (\ref{FROMTHE3}) implies that $\beta^2=-1$, or that $\beta=\pm{i}$.\par 
\indent 
The final result is
\begin{eqnarray}
\label{SELFISH4}
F^a_{0i}=\bigl(\epsilon_{ijk}N^k+\beta\underline{N}h_{ij}\bigr)B^j_a
\end{eqnarray}
\noindent 
where $\beta=\pm{i}$, which is the 3+1 decomposition of the Hodge duality condition of the GYM field strength $F^a_{\mu\nu}$ with respect to the metric $g_{\mu\nu}$ which it couples to.\par 
\indent 
The results of this appendix are based on the Lorentzian signature case.  To extend these results to spacetimes of Euclidean signature, one needs only to perform a Wick rotation $N\rightarrow-iN$, and all the analogous results of this paper carry through.

\end{document}